\documentclass[12pt,preprint,nofootinbib,noshowkeys,noshowpacs,longbibliography,
superscriptaddress,tightenlines,floatfix]{revtex4-2}
\usepackage{amsmath,amsfonts,amsthm,amssymb}
\usepackage{graphics,graphicx}
\usepackage{color}
\definecolor{darkblue}{RGB}{0,0,196}
\definecolor{darkgreen}{RGB}{0,120,0}

\def\beq{\begin{equation}}
\def\eeq{\end{equation}}
\def\st{\begin{equation}}
\def\stp{\end{equation}}
\def\ba{\begin{eqnarray}}
\def\ea{\end{eqnarray}}

\usepackage[colorlinks=true,linktocpage=true,linkcolor=darkblue,citecolor=red,urlcolor=darkblue]{hyperref}

\begin{document}
\preprint{}        
\title{Some insights on the partonic collectivity in heavy-ion collisions}
 \author{Rajeev Singh}
    \email{rajeev.singh@e-uvt.ro}
    \affiliation{Department of Physics, West University of Timisoara, Bd.Vasile P\^arvan 4, Timisoara 300223, Romania}
\date{\today}
	\bigskip
\begin{abstract}
Elliptic flow and Number of Constituent Quarks scaling provide crucial insights into the underlying dynamics and degrees of freedom in heavy-ion collisions. This article provides some insights on the transition from hadronic to partonic collectivity, revealing possible key signatures of medium evolution. At low energies around ($\sqrt{s_{\rm NN}} \leq 4.0$ GeV), $v_2$ may be negative and NCQ scaling might break down. We expect that as the collision energy rises beyond 4.0 GeV, $v_2$ may become positive and NCQ scaling gradually restores. The transition in $v_2$ behavior, coupled with the breakdown and restoration of NCQ scaling, highlights the increasing significance of partonic interactions at higher energies, marking the onset of partonic collectivity and the emergence of quark-gluon plasma (QGP)-like properties.
%
\end{abstract}
\maketitle
\newpage
\section{Introduction}
\label{sec:intro}
Partonic collectivity serves as a crucial indicator for the formation of QGP in high-energy nuclear collisions. A key manifestation of this collectivity is the scaling of elliptic flow, $v_2$, with the number of constituent quarks (NCQ), which has been prominently observed in light hadrons produced in top-energy nuclear collisions at the Relativistic Heavy Ion Collider (RHIC) and the Large Hadron Collider (LHC)~\cite{Arslandok:2023utm}. This scaling behavior is a hallmark of partonic dynamics. 
Interestingly, NCQ scaling is expected to break down at much lower collision energies, such as 3--4 GeV.
This may be signaling a transition to a hadronic-interaction-dominated equation of state. This transition may suggests that the system's dynamics are primarily governed by hadronic rather than partonic interactions at these energies. 
However, it is also expected that as the collision energy increases beyond 4 GeV in Au+Au collisions, a gradual reappearance of NCQ scaling may be observed, indicating the resurgence of partonic interactions. We suspect that if the breakdown and subsequent restoration of NCQ scaling is observed then this will provide compelling evidence for the gradual onset of partonic collectivity in the intermediate-energy nuclear collisions indicating the complex interplay between hadronic and partonic phases in the evolution of the system. These expectations remain subject to experimental verification and are, at present, theoretical conjectures grounded in the framework of partonic physics~\cite{Dunlop:2011cf,Du:2023ype,Hatwar:2024khh,Zhu:2025kud}.
\section{Elliptic flow and NCQ scaling}
Elliptic flow ($v_2$), defined as the second-order harmonic coefficient in the Fourier expansion of the azimuthal distribution of final-state particles relative to the reaction plane, serves as a key observable for studying the interactions and degrees of freedom in the matter produced during heavy-ion collisions~\cite{Bhalerao:2005mm,Voloshin:2008dg,Dusling:2010rm,Epelbaum:2014ika}. This quantity is highly sensitive to the underlying dynamics of the system, particularly the constituent interactions and the state of the medium. 
A substantial $v_2$ signal, along with the scaling of $v_2$ with the NCQ, is widely regarded as strong evidence for the formation of QGP~\cite{Hwa:2002tu,PHENIX:2004vcz,STAR:2005gfr,Jia:2006vj,Braun-Munzinger:2007edi,Heinz:2013th,STAR:2015gge,STAR:2015vvs,STAR:2016ydv,STAR:2017kkh,STAR:2020dav}. 
The phenomenon of NCQ scaling refers to the empirical observation that the elliptic flow of various particles collapses onto a universal curve when normalized by NCQ. This behavior is indicative of the presence of quark degrees of freedom in the medium and reflects the collective motion at the partonic level.

However, as the collision energy decreases, the conditions of high temperature and energy density required for QGP formation are no longer met. Below a certain energy threshold, signals such as significant $v_2$ and NCQ scaling are expected to vanish, marking the transition from a partonic to a hadronic-dominated system. To investigate this transition and explore the phase structure of QCD, the Beam Energy Scan (BES) program at RHIC systematically reduces the collision energy. The BES program spans a broad energy range ($\sqrt{s_{\rm NN}} = 3$ to 62.4 GeV) and aims to search for potential signatures of a QCD first-order phase boundary and the critical point through heavy-ion collision experiments~\cite{Fukushima:2010bq,Bzdak:2019pkr,Luo:2020pef}.

During the RHIC BES-I phase, $v_2$ measurements revealed that NCQ scaling was relatively well maintained in collisions with $\sqrt{s_{\rm NN}} \geq 7.7$ GeV~\cite{PHENIX:2006dpn,PHENIX:2012swz,STAR:2012och,STAR:2013cow,STAR:2013ayu,STAR:2015rxv}. This scaling behavior suggests the dominance of partonic collectivity in the medium at these energies. However, deviations from NCQ scaling were observed, particularly for the $\phi$-meson $v_2$, where a discrepancy of approximately $2\sigma$ was noted in collisions at $\sqrt{s_{\rm NN}} = 7.7$ and 11.5 GeV~\cite{STAR:2012och,STAR:2013cow,STAR:2013ayu,STAR:2015rxv}. These findings indicate that the behavior of $\phi$-meson $v_2$ may not fully conform to the universal NCQ scaling trend at these energies, warranting further investigation with larger and more statistically robust data samples.
Recent results from the STAR experiment at even lower collision energy, specifically at $\sqrt{s_{\rm NN}} = 3$ GeV, show a clear breakdown of NCQ scaling among $\pi^+$, $K^+$, and proton $v_2$~\cite{STAR:2021yiu}. This breakdown is indicative of a transition from partonic to hadronic dynamics as the collision energy decreases, marking the emergence of a hadronic-dominated regime.

The RHIC BES-II phase has been designed to extend these investigations over a collision energy range of $\sqrt{s_{\rm NN}} = 3$ to 19.6 GeV~\cite{Cleymans:2005xv,Andronic:2005yp,STAR:2017sal}. This expanded energy range aims to provide a deeper understanding of the QCD phase structure and the interplay between partonic and hadronic degrees of freedom in the evolution of the collision system.
At low collision energies, particles are preferentially emitted orthogonal to the reaction plane. This may result in a negative $v_2$ signal. However, as the collision energy increases, $v_2$ as a function of transverse momentum ($p_T$) is expected to become positive. This expected shift indicates a rapid reduction in the spectator-shadowing effect within this energy range.
At lower collision energies, it is also expected that the Multi-Phase Transport Model with Hadron Cascade (AMPT-HC)~\cite{Lin:2004en} may be able to describe qualitatively the measured $v_2$ data. However, we must keep in mind that $v_2$ might be sensitive to different modeling assumptions, particularly regarding the role of partonic versus hadronic interactions, suggesting that partonic interactions should play a role in generating the significant $v_2$ signal at higher energies, awaiting confirmation from the experiments. Furthermore, the scaling of $v_2$ with NCQ is expected to provide insight into the effective degrees of freedom in the medium. The presence or absence of NCQ scaling serves as a key indicator of whether the system is dominated by partonic or hadronic interactions, offering a window into the nature of the medium created in these collisions.

As the collision energy increases, the degree of NCQ scaling should gradually improve. This expected trend suggests that at lower energies, such as $< 4.0$ GeV, the equation of state of the created matter is predominantly governed by hadronic interactions. In contrast, partonic interactions play an increasingly significant role at higher collision energies marking the onset of partonic collectivity.
From a modeling perspective, the AMPT model might capture the physics of NCQ scaling.
It is expected that the $p_T$-integrated $v_2$ may change its sign going from lower to higher collision energies. This behavior aligns with the mechanism of baryon number transport, where quarks transported from beam rapidity to mid-rapidity undergo more intense scatterings compared to quarks produced directly at mid-rapidity. 
Furthermore, the initial nuclear matter is neutron-rich, which enhances the transported effect for antiparticles relative to particles. This asymmetry reflects the influence of the neutron-rich environment on particle dynamics, providing additional insight into the interplay between baryon transport and the medium's evolving state across different collision energies.

\section{Outlook}



The expected breakdown of NCQ scaling at lower energies and its subsequent restoration at higher energies may highlight the transition from hadronic to partonic dominance in the system. This progression underscores the increasing importance of partonic interactions, signaling the onset of partonic collectivity and the emergence of quark-gluon plasma-like properties in the created matter~\cite{Arslandok:2023utm}.
\section*{Acknowledgments}
R.S. is supported by a postdoctoral fellowship of West University of Timișoara, Romania. This article came up from the discussions with the participants in the XXVI DAE-BRNS High Energy Physics Symposium 2024.
\bibliographystyle{utphys}
\bibliography{pv_ref}
\end{document}